\documentstyle[aps,pra,multicol,epsfig,amstex]{revtex}

\newcommand{\ve}[1]{{\mathbf{#1}}}
\newcommand{\mat}[1]{\Bar{\Bar{#1}}}
\begin{document}
\title{Collective Modes of Trapped Fermi Gases in the Normal Phase}

\author{G.\ M.\ Bruun}
\address{Nordita, Blegdamsvej 17, 2100 Copenhagen, Denmark}
\maketitle

\begin{abstract} 
We consider the collective mode spectrum of a normal Fermi gas in a spherical harmonic trap. 
Using a self-consistent random-phase-approximation, we systematically examine the effects of the two-body
 interactions on the modes of various symmetries.  For weak coupling where the spectrum is 
shifted very little away from the ideal gas case, a sum-rule approach is shown to work well.
 For stronger coupling,  the interplay between the single particle and the collective excitations 
causes effects such as mode splitting and Landau damping. A finite low frequency response present at 
stronger coupling is predicted for the quadrupole mode. Finally, we briefly discuss the effect of a finite
 temperature on the spectrum and the excitation of higher collective modes.
\end{abstract}

Pacs Numbers: 32.80.Pj, 05.30.Fk, 67.55.Jd

\section{Introduction}
Trapped Fermi gases is rapidly emerging as a new important area of research in the field of ultracold atomic gases.
Impressive experimental results have been presented with the demonstration of trapping and 
cooling of fermionic alkalis  using a 
magnetic trap for  $^{40}$K~\cite{DeMarcoScience} and $^6$Li~\cite{Schreck},
 and  an optical trap for $^6$Li~\cite{O'Hara}. These systems provide 
a large degree of experimental control: One can manipulate the effective particle-particle interaction by tuning 
an external magnetic field~\cite{Houbiers}, the geometry of the system can be changed, the number of internal
 degrees of freedom of the constituent particles is given by  the number of trapped hyperfine states, and the fundamental
 quantum statistics can be altered by trapping alkali atoms of Bose, Fermi (or both) statistics. Ultracold atomic Fermi gases  
therefore provide 
a very rich system to study with useful  concepts coming from atomic, condensed matter, nuclear, and statistical physics
and  several theoretical papers have already been published focusing on various aspects of this 
subject~\cite{Normal,Bruun1,BCS}.

 The high precision spectroscopy of collective modes possible in these systems 
combined with the ability to manipulate the interaction strength makes the study of collective modes 
for trapped Fermi gases  a promising subject. Indeed, such studies have proven to be very useful in the 
field of Bose-Einstein condensation~\cite{Edwards1}. The collective modes which can be observed as shape 
oscillations of the atomic cloud 
correspond to sound waves in a homogeneous gas, but for a confined system their spectrum is discrete. 
It is well known that there in general are two regimes in which to study these modes~\cite{Pines}:
When the lifetime $\tau$ of the quasiparticles is much longer
than the characteristic period of motion ({\it{i.e.}} $\omega_T\tau\gg 1$
for atoms in a harmonic trap of frequency $\omega_T$), 
there are few scattering events per sound oscillation, and the restoring
force is due to the self-consistent mean field of the gas. 
Wave motion encountered in this limit is designated ''zero sound''. 
For the hydrodynamic regime $\omega_T\tau\ll 1$, on the other hand, collisions
ensure local thermodynamic equilibrium. Hydrodynamic sounds waves are sometimes called ``first sound''.
Due to the large variations in the effective interaction between the relevant alkali 
atoms in different hyperfine states, both regimes 
could be experimentally relevant and they have thus both been studied theoretically~\cite{Vichi,BruunPRL}. 

We will in this paper examine the spectrum of collective modes in the collisionless regime for a gas in a
spherical harmonic 
trap. By using a formalism well-known from nuclear physics, we systematically study the effect of the interactions on various 
 aspects of the collective mode spectrum for modes of 
monopole, dipole, and quadrupole symmetry when the gas is in the normal phase. We show that for weak coupling,
 the spectrum is characterized by a large degree of collectivity (giant resonances). For such weak interactions where the 
shift of the collective mode spectrum away from the ideal gas result is very small, a sum rule calculation of the 
collective mode spectrum assuming a non-interacting ground state is shown to work well. With increasing coupling however,
the effect of the interactions on the ground state become significant thus complicating a sum rule calculation. 
For stronger coupling, we demonstrate that effects such as Landau damping and mode splitting 
 become important for some modes. The magnitude of these effects, which depends on the overlap between the single particle
 and the collective response,  is shown to be different for the various modes considered. 
Finally,  we briefly discuss the effects of a finite temperature on the spectrum, and higher modes relevant 
for the thermodynamics of the system are considered. 

\section{Formalism}
In this section, we outline the basic theoretical framework used to
calculate the collective mode spectrum of a trapped gas of interacting 
fermions. We consider a gas of fermionic atoms of mass $m$, confined by a
  potential $U_0(\ve{r})$, with an equal number of atoms $N_\sigma$ in each
of two hyperfine states, $|\sigma=\uparrow,\downarrow\rangle$. Two fermions in the same 
internal state $\sigma$ must have odd relative orbital angular momentum 
(minimally $p$-wave), and at
 low temperatures $T$ the centrifugal barrier suppresses their mutual interaction~\cite{DeMarco}.
Thus, we assume the interaction to be effective only between atoms
in different hyperfine states, and to be dominated by
the low-energy $s$-wave scattering characterized by the scattering length $a$.  
Within mean field (Hartree Fock) theory,  the gas is then described by the Hamiltonian:
\begin{gather}\label{Hamiltonian}
\hat{H}=\sum_{\sigma}\int d^3r\,
 \psi_{\sigma}^{\dagger}({\mathbf{r}})
[ \frac{-\hbar^2}{2m}\nabla^2+\frac{1}{2}m\omega_T^2r^2-\mu_F]\psi _{\sigma}({\mathbf{r}}) 
+g\sum_\sigma\int d^3r\,  \langle\psi_\sigma^{\dagger}({\mathbf{r}})\psi_\sigma({\mathbf{r}})\rangle
\psi_{\bar{\sigma}}^{\dagger}({\mathbf{r}})\psi_{\bar{\sigma}}({\mathbf{r}})+\hat{F}(t),
\end{gather}
where  $\langle\ldots\rangle$ denotes the thermal average, and the field operators  
$\psi_{\sigma}({\mathbf{r}})$ obey the usual fermionic anticommutation rules describing
 the annihilation of a fermion at position ${\mathbf{r}}$ in the hyperfine 
state $|\sigma\rangle$.  The trapping 
potential is for simplicity assumed to be an isotropic harmonic oscillator 
 $U_0(\ve{r})=\frac{1}{2}m\omega_T^2 r^2$, $\mu_F$ is the chemical potential,  
and $g=4\pi a\hbar^2/m$.
 $\hat{F}(t)$ is an external perturbation with which we probe the system. The subject of this paper is
the effects of the interaction on the collective mode spectrum of  the gas in the normal phase  and we are therefore 
 neglecting pairing terms describing a possible superfluid transition.

When the atom-atom interaction is rather weak such that the 
life-time $\tau$ of the quasiparticles is much larger than $1/\omega_T$, the appropriate 
way to calculate the spectrum is the self-consistent random phase
 approximation (RPA)~\cite{Bohm}. The problem of 
collective modes of a  spatially confined two-component Fermi system has been studied extensively in the 
field of nuclear physics~\cite{Bohr}. Thus, the basic formalism used in this paper is well-known 
from this field.

  As we are interested in 
modes corresponding to density fluctuations excited by  perturbations of the kind 
 $\hat{F}(t)\propto\exp(i\omega t)\sum_\sigma\int d^3rF_\sigma(\ve{r})\hat{\rho}_\sigma(\ve{r})$,
 we consider the retarded density-density correlation function 
$\langle\langle\hat{\rho}_\sigma(\ve{r})\hat{\rho}_{\sigma'}(\ve{r}')\rangle\rangle(\omega)$.
Here, $\hat{\rho}_\sigma(\ve{r})=\psi^\dagger_\sigma(\ve{r})\psi_\sigma(\ve{r})$
is the density operator for atoms in the hyperfine state $|\sigma\rangle$ and 
  $\langle\langle AB\rangle\rangle(\omega)$ is the Fourier transform of the 
retarded function  $-i\Theta(t-t')\langle[A,B]\rangle$~\cite{Fetter}. 
Within RPA, we have 
\begin{equation}\label{Inversion}
\mat{\Pi}(\omega) =[1-\mat{g}\mat{\Pi}_0(\omega)]^{-1}\mat{\Pi}_0(\omega)
\end{equation}
with 
\begin{equation}
\mat{\Pi}(\omega) =\left\{
\begin{array}{cc}
\langle\langle\hat{\rho}_\uparrow(\ve{r})\hat{\rho}_\uparrow(\ve{r}')\rangle\rangle(\omega) &
\langle\langle\hat{\rho}_\uparrow(\ve{r})\hat{\rho}_\downarrow(\ve{r}')\rangle\rangle(\omega)\\
\langle\langle\hat{\rho}_\downarrow(\ve{r})\hat{\rho}_\uparrow(\ve{r}')\rangle\rangle(\omega) &
\langle\langle\hat{\rho}_\downarrow(\ve{r})\hat{\rho}_\downarrow(\ve{r}')\rangle\rangle(\omega)
\end{array}
\right\},
\end{equation}
\begin{equation}\label{Pi0}
\mat{\Pi}_0(\omega) =\left\{
\begin{array}{cc}
\langle\langle\hat{\rho}_\uparrow(\ve{r})\hat{\rho}_\uparrow(\ve{r}')\rangle\rangle_0(\omega) &0\\
0 &\langle\langle\hat{\rho}_\downarrow(\ve{r})\hat{\rho}_\downarrow(\ve{r}')\rangle\rangle_0(\omega)
\end{array}
\right\}
\end{equation}
and 
\begin{equation}
\mat{g}=\frac{g}{\hbar}\left\{
\begin{array}{cc} 0&\delta(\ve{r}-\ve{r}')\\ \delta(\ve{r}-\ve{r}')&0
\end{array}
\right\}.
\end{equation}
Here the matrix products denote integrals:
 $\mat{A}\mat{B}\equiv\int d^3r''A(\ve{r},\ve{r}'')B(\ve{r}'',\ve{r}')$. The independent 
particle correlation functions 
 $\langle\langle\ldots\rangle\rangle_0$ are calculated within the time-independent 
 Hartree-Fock (HF) approximation.
 In order to describe the effect of the interactions on the collective mode spectrum correctly,
it is crucial that one uses the self-consistent static HF states to
calculate the independent particle correlation functions $\langle\langle\ldots\rangle\rangle_0$.
 If one simply uses the non-interacting trap states, the effect of the interactions 
on the single particle excitations are ignored and the predicted collective mode spectrum from 
Eq.~(\ref{Inversion}) will be incorrect.

For a spherically symmetric trap, the correlation functions split into terms describing the 
various multipole modes. Indeed, we have
\begin{gather}\label{Angular}
\langle\langle\hat{\rho}_\sigma(\ve{r})\hat{\rho}_{\sigma'}(\ve{r}')\rangle\rangle(\omega)=
\sum_{LM}\rho\rho(r\sigma,r'\sigma',\omega)_LY_{LM}(\theta,\phi)Y_{LM}(\theta',\phi')^*
\end{gather}
with $r$ denoting the radial distance to the center of the trap and the $Y_{LM}(\theta,\phi)$ being
 the usual spherical harmonics. For the independent particle correlation functions,
 we obtain for the radial part~\cite{Bertsch}:
\begin{gather}\label{Radial}
\rho\rho_0(r\sigma,r'\sigma',\omega)_L=
\sum_{\eta\eta'll'}\frac{\delta_{\sigma,\sigma'}\hbar
(2l+1)(2l'+1)}{4\pi(2L+1)r^2{r'}^2}|\langle ll'00|L0\rangle|^2
\frac{u_{\eta l}(r)u_{\eta l}(r')u_{\eta'l'}(r)u_{\eta'l'}(r')}{\xi_{\eta l}-\xi_{\eta'l'}-\hbar\omega-i\delta}
(f_{\eta l}-f_{\eta'l'}).
\end{gather}
The quasiparticle wave functions $u_{\eta l}(r)$ are obtained from a self-consistent solution to 
mean field Hamiltonian in Eq.(\ref{Hamiltonian}) with $\hat{F}(t)=0$ writing the HF wave 
functions on the form  $u_{\eta lm}(\ve{r})=u_{\eta l}Y_{LM}(\theta,\phi)/r$~\cite{Bruun1}. The quasiparticles
 $u_{\eta lm}(\ve{r})$ have energies $\xi_{\eta l}$ and $f_{\eta l}=[\exp(\beta\xi_{\eta l})+1]^{-1}$
is the usual Fermi function with $\beta=1/k_BT$. We write $\langle ll'00|L0\rangle$ for the Clebsch-Gordan coefficients.

The structure of the self-consistent RPA calculation is then the usual one: First we obtain a self-consistent solution 
to the mean field Hamiltonian in Eq.\ (\ref{Hamiltonian}) with $\hat{F}=0$; then 
 $\mat{\Pi}_0(\omega)$ is formed from Eq.(\ref{Pi0}),(\ref{Angular})-(\ref{Radial}) for varying external frequency 
 $\omega$; and finally the RPA response function is calculated from Eq.\ (\ref{Inversion}). 
 The poles of $\mat{\Pi}(\omega)$ then give the collective modes of the gas excited by the density operator. 

It follows from Eq.\ (\ref{Angular}) that the spherical 
symmetry permits an independent calculation of the radial part of the response functions for each spherical 
multipole, and the numerical dimension of the matrices is therefore determined by the number of radial mesh 
points. Hence, the method outlined is computationally very efficient as compared to a usual particle-hole 
representation of the RPA  which in general yields matrices of very large dimensions~\cite{Blaizot}. 

\section{Observables and Sum Rules}
A quantity of great physical interest is the strength function directly related to  the net transitions per unit time with
 energy $\hbar\omega$ induced by the operator $\hat{F}$: 
\begin{equation}
S(F,\omega)\equiv \sum_{nm}\frac{e^{-\beta E_n}-e^{-\beta E_m}}{\mathcal{Z}}|\langle n|\hat{F}|m\rangle|^2
\delta(\hbar\omega+E_n-E_m).
\end{equation}
Here ${\mathcal{Z}}$ is the grand partition function and $|n\rangle$ is an eigenstate of the Hamiltonian with energy $E_n$.
 For operators of the form $\hat{F}(t)\propto\sum_\sigma\int d^3rF_\sigma(\ve{r})\hat{\rho}_\sigma(\ve{r})$,
it can be obtained as 
\begin{equation}\label{strength}
S(F,\omega)=-\frac{1}{\hbar\pi}\sum_{\sigma,\sigma'}\int d^3rd^3r'F_\sigma(\ve{r})F_{\sigma'}(\ve{r}')
Im[\Pi_{\sigma,\sigma'}(\ve{r},\ve{r}',\omega)].
\end{equation}
The strength function is the main object of the present paper. We will calculate it for perturbing 
operators of the form $F_\sigma(\ve{r})=F_\sigma(r)Y_{LM}(\theta,\pi)$ exciting the monopole ($L=0$), 
dipole ($L=1$), and quadrupole ($L=2$) modes. We take $F_\uparrow(r)=F_\downarrow(r)$ for the monopole 
and quadrupole modes thereby exciting total density fluctuations of the given symmetries. For the  dipole
symmetry, we  take $F_\uparrow(r)=\pm F_\downarrow(r)$ exciting both the dipole, and the spin-dipole mode. 
For the dipole mode, we have  $\omega=\omega_T$ independent of the interactions since it simply
 corresponds to a center-of-mass
oscillation of the cloud. We calculate it to check the consistency of our numerics.

If one defines the moments $m_k\equiv\int d\omega S(F,\omega)\omega^k$, the frequency $\tilde{\omega}$
 of the mode excited by  $\hat{F}$ can be calculated as 
\begin{equation}\label{sumrule}
\tilde{\omega}=\sqrt{\frac{m_{k+2}}{m_k}}
\end{equation}
 if the operator only excites one mode (giant resonance). This method is very handy  since the two 
moments can be calculated as  $m_1=\langle0|[\hat{F},[\hat{H},\hat{F}]]|0\rangle$ and 
 $m_3=\langle0|[[\hat{F},\hat{H}],[\hat{H},[\hat{H},\hat{F}]]]|0\rangle$; i.e.\ one only needs information about the 
ground state of the system to calculate the excitation energy~\cite{Lipparini}. 

An important rule coming from particle conservation is 
the $f$-sum rule~\cite{Bertsch}:
\begin{equation}
m_1=\int d\omega S(F,\omega)\omega=\frac{1}{m}\sum_\sigma\int d^3r|\nabla F_\sigma(\ve{r})|^2
\rho_\sigma(\ve{r})
\end{equation}
with $\rho_\sigma(\ve{r})=\langle\hat{\rho}_\sigma(\ve{r})\rangle$ being the equilibrium density. 
This identity provides another  important check on the reliability of our RPA calculations as the 
equilibrium density is easily obtained from the static HF solution. For our numerical RPA results, this 
sum rule turns out to be obeyed to within $\sim3\%$ indicating the accuracy of our calculations.

\section{Results}
In this section, we present some typical results concerning the effect of the interactions on the collective mode spectrum 
obtained by solving the RPA equations for varying coupling 
strengths.

  As outlined above, if the perturbing operator only excites one state, then a sum-rule calculation  
based on Eq.\ (\ref{sumrule}) should give the correct energy of the excited state if an accurate ground state 
is used. Based on this assumption and on the fact that for weak interactions
the ground state can be well approximated by an ideal Fermi gas, the lowest collective mode of 
monopole, (spin)-dipole, and quadrupole symmetry was calculated in Ref.~\cite{Vichi}.
 Several  effects can make such a calculation 
invalid: The interactions can change the ground state significantly away from the ideal gas limit although the 
operator $\hat{F}$ still excites a  single collective mode (giant resonance). For stronger couplings, the 
mode can be  fragmented into several modes and/or an incoherent excitation of several single
 particle states. This latter effect is known as Landau damping.
 We will  now consider the importance of these  effects for varying coupling strengths.

\begin{figure}
\centering
\epsfig{file=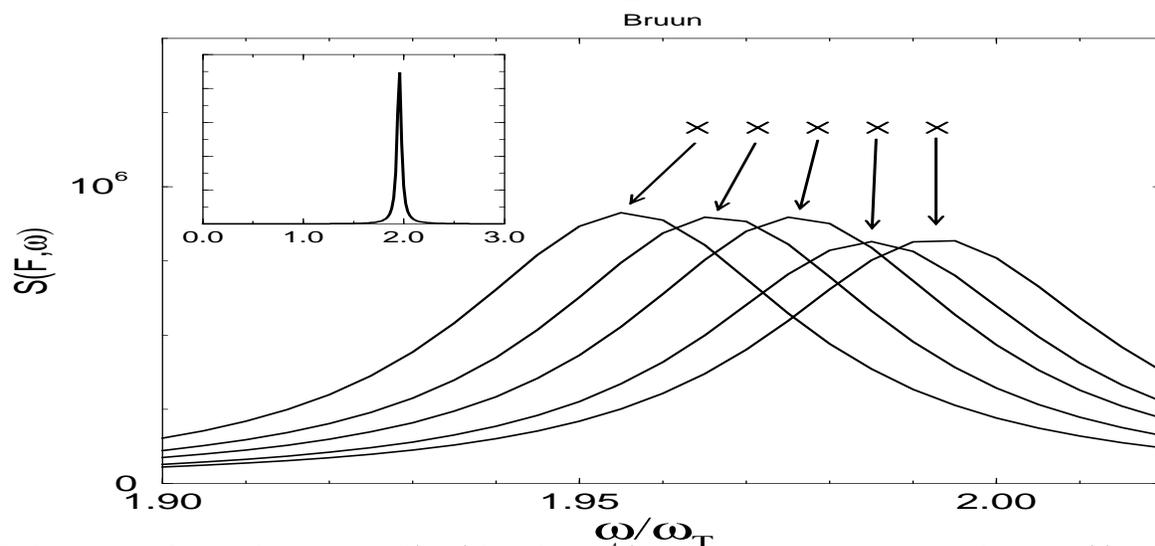,height=\textwidth,width=0.34\textheight,angle=-90}
\caption{The $T=0$ collective response $S(F,\omega)$ in units of $l_h^4/\hbar\omega_T$ 
for the monopole mode excited by $F(r)=r^2$ for various coupling strengths.
 The sum-rule results are connected with the corresponding RPA results by an arrow. The inset shows $S(F,\omega)$
for $k_F|a|\simeq0.32$ over a larger frequency range.}
\label{sumrulefig}
\end{figure}

For very weak interactions, we expect the sum-rule approach to give an accurate description of the lowest 
collective modes.  We have checked this by comparing our RPA results for small coupling strengths with the 
prediction based on a sum rule calculation. For simplicity, we here only show results for the monopole mode 
 [$F_\sigma(\ve{r})=F(r)Y_{00}(\theta,\phi)$].  Completely equivalent results are obtained for the spin-dipole and the 
quadrupole modes. Using a local density approximation for the $T=0$ ground state of an ideal 
gas, a sum rule calculation predicts the frequency of the lowest monopole mode  to 
be $w_M\simeq2\omega_T(1+0.3\times\frac{3}{8}k_Fa)^{1/2}$  with $k_F=\sqrt{2m\mu_F/\hbar^2}$ being the Fermi 
vector in the center of the trap~\cite{Vichi}. 
Figure (\ref{sumrulefig}) depicts  a typical result obtained from a RPA calculation. 
We plot $S(F,\omega)$ (in units of $l_h^4/\hbar\omega_T$ with $l_h=\sqrt{\hbar/m\omega_T}$ being the trap length)
 given by Eq.\ (\ref{strength}) 
for $F_\sigma(r)=r^2$. The parameters are chosen such that there are $\sim 10^4$ atoms trapped and $T=0$. Also, 
we have added a small imaginary part $\Gamma=0.05\hbar\omega_T$ to the frequency to model a smooth response 
typically observed experimentally. The strength of the interaction is characterized by the parameter   
 $k_Fa$. 
The response is calculated as a function of the frequency $\omega$ for the coupling strengths: 
 $k_F|a|\simeq 0.06, 0.13, 0.19, 0.25$ and $0.32$. We have chosen an attractive interaction ($a<0$) such that the 
frequency shift away from the ideal gas result $\omega_M=2\omega_T$ is increasingly negative with increasing $|a|$.
  The prediction for $\omega_M$ based on the sum-rule 
approach is indicated by the symbol $\times$ for the various coupling strengths.

As can be seen from Fig.\ (\ref{sumrulefig}), there is good agreement between the prediction for the frequency 
of the mode based on the sum-rule 
approach and the RPA result for $k_F|a|\lesssim 0.2$. For such weak coupling strengths, the $r^2$ perturbation 
only excites the lowest monopole mode and the ground state is essentially given by the ideal gas result. For
 stronger interactions, the sum-rule result underestimates the frequency shift of the
 monopole mode due to the interactions. 
As can be seen from the inset in fig.\ (\ref{sumrulefig}), the $r^2$ still essentially 
only excites one mode (giant resonance) for these coupling strengths. This means that a sum-rule calculation in 
principle should yield the correct result if one used to the exact ground state.  We conclude that the discrepancy in 
fig.\ (\ref{sumrulefig}) is due to the fact that the simple non-interacting approximation
 to the ground state is inadequate in a sum-rule calculation for these coupling strengths.

We have performed calculations with varying number of particles trapped and different coupling strengths for both the  
monopole, spin-dipole, and the quadrupole modes and compared with the predictions based on the sum-rule 
approach as given in Ref.~\cite{Vichi}. The general conclusion is, that the simple sum-rule results  
work well for  $k_F|a|\lesssim 0.2$
where  the frequency shift away from the ideal gas result is very small for all these modes. 
For stronger interactions, one still has a giant resonance for the lowest monopole, spin-dipole, and 
quadrupole modes. However, to get an accurate prediction for the eigenfrequencies for these coupling
strengths based on a sum-rule calculation, one would need to take into account the effect of the interactions on the 
ground state. Such a calculation could involve determining the ground state using a static Hartree Fock 
theory and then calculating the moments $m_k$. 

\begin{figure}
\epsfig{file=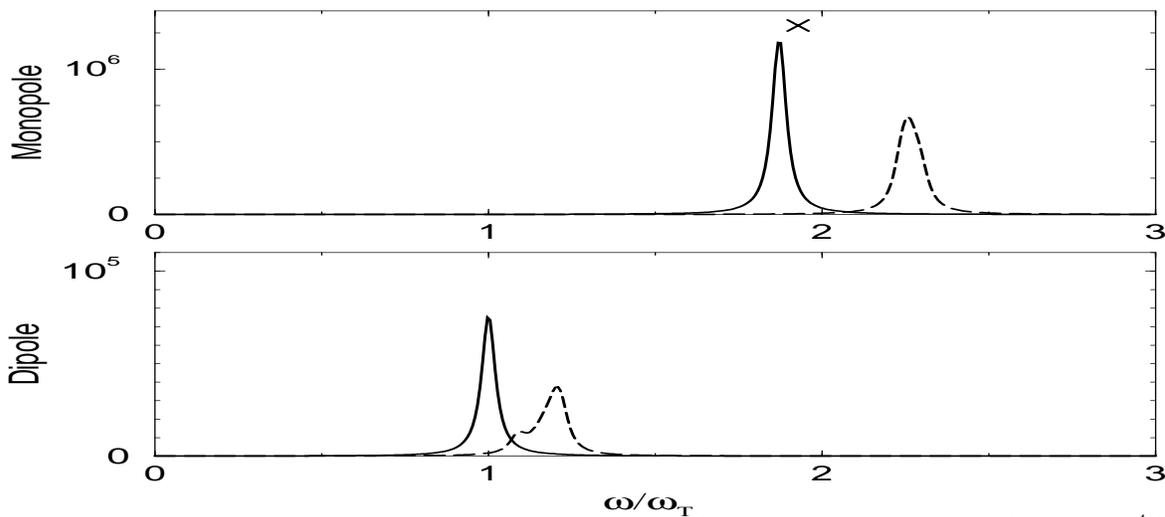,height=\textwidth,width=0.34\textheight,angle=-90}
\caption{The $T=0$ collective (solid line) and single particle (dashed line) response for the monopole (in units of 
	$l_h^4/\hbar\omega_T$) and dipole mode (in units of $l_h^2/\hbar\omega_T$) excited by 
 $F_\sigma(r)=r^2$  and $F_\sigma(r)=r$ respectively. The coupling strength is $k_F|a|\simeq0.6$.}
\label{MonoDifig}
\end{figure}

\begin{figure}
\epsfig{file=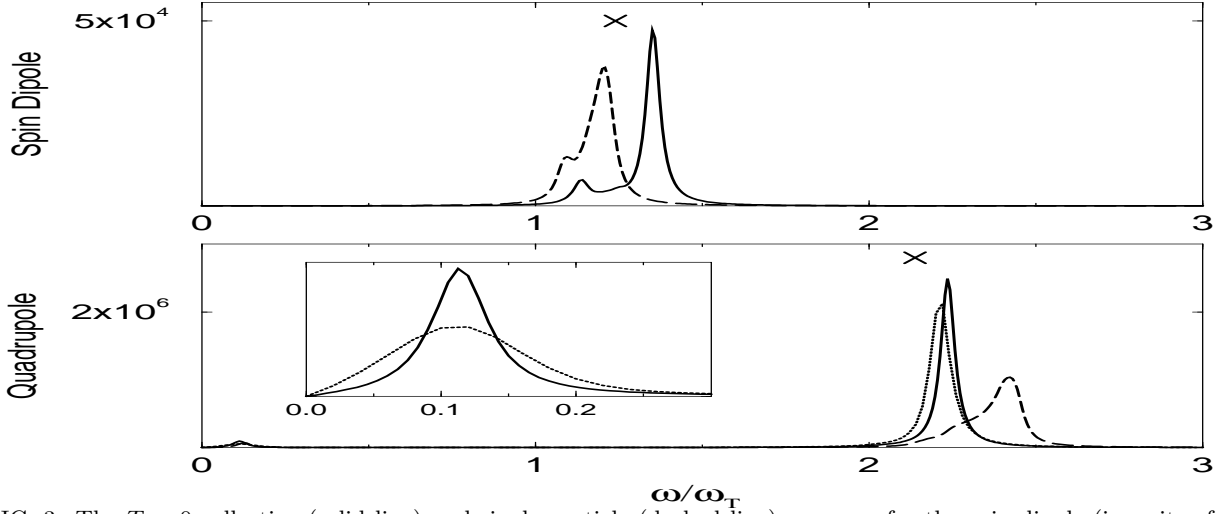,height=\textwidth,width=0.34\textheight,angle=-90}
\caption{The $T=0$ collective (solid line) and single particle (dashed line) response for the spin-dipole 
(in units of $l_h^2/\hbar\omega_T$) and the quadrupole mode (in units of $l_h^4/\hbar\omega_T$) excited by 
 $F_\sigma(r)=-F_{\bar{\sigma}}(r)=r$ and $F_\sigma(r)=r^2$ respectively. The coupling strength is $k_F|a|\simeq0.6$.
For the quadrupole mode, the dotted line depicts the collective response for $k_BT=3\hbar\omega_T$. 
The inset shows in more detail the low frequency collective quadrupole response.
}
\label{SpinDiQuadrufig}
\end{figure}

We now consider the collective mode spectrum for more strongly interacting systems where the eigenmodes differ 
significantly from the ideal gas result. In this regime, the results based on the simple sum-rule 
calculation are invalid. This limit is especially relevant for experiments on $^6$Li, where large an negative 
scattering lengths between certain hyperfine states have been predicted theoretically~\cite{Houbiers} and recently
measured experimentally~\cite{O'Hara}. In fig.\ (\ref{MonoDifig})-(\ref{SpinDiQuadrufig}), we plot 
the response $S(F,\omega)$ for the monopole, spin-dipole, 
and quadrupole symmetries excited by  $F_\sigma(r)=r^2$, $F_\sigma=-F_{\bar{\sigma}}=r$, and $F_\sigma(r)=r^2$ respectively. 
We also plot the dipole response excited by $F_\sigma=r$ to further establish the accuracy of the calculations. 
There are $\sim 1.6\times10^4$ particles trapped, $k_Fa\simeq-0.6$, $\Gamma=0.05\hbar\omega$, and $T=0$. 
 The symbols $\times$ again indicate the sum-rule predictions for
the collective modes~\cite{Vichi}. We also plot the single particle response $S_0(F,\omega)$ based on using 
 $\Pi_0(\omega)$ instead of  $\Pi(\omega)$ in Eq.(~\ref{strength}). The collective response is indicated by solid lines
 whereas the single particle response is indicated by dotted lines. 

 First, we note from fig.\ (\ref{MonoDifig}) that 
the dipole response is very accurately peaked at $\omega=\omega_T$ as it should be. This provides another confirmation 
of the accuracy of our calculations. Also, we see that the single particle response (dashed lines) in general is 
moved to higher frequencies as compared to the ideal gas result for all modes. This is because the 
mean field is attractive for $a<0$ and concentrated in the center of the trap. Since the spatial extend of the trap 
states in general increases with their energy, the decrease in the corresponding eigen-energies due to the mean field
 is larger for the  lower trap states compared to for the higher states. The gap between the harmonic
 oscillator bands thus increases compared to the non-interacting limit resulting in a shift of the single particle response 
to higher frequencies~\cite{Bruun1,BruunDetect}. Correspondingly, for a repulsive interaction
 the shift in the single particle response would have been to lower frequencies. 
We see that for all modes (except, of course, for the dipole mode),
  there is a significant disagreement between the sum-rule results and the RPA prediction as expected.

Figure (\ref{MonoDifig}) indicates that the monopole mode is still characterized by a  single giant resonance
 excited by the operator $r^2$. This is because contrary to the spin-dipole and quadrupole modes, the 
collective response for the monopole mode is moved to lower frequencies compared to the ideal gas result 
($\omega_M=2\omega_T$) due to the 
interactions. As can be seen from fig.\ (\ref{MonoDifig}), the overlap between the collective response and the single
 particle response (which is moved to frequencies $\omega\gtrsim2\omega_T$) is therefore 
insignificant prohibiting a decay of the collective mode into single particle excitations.  The monopole  mode is 
therefore well-defined.

The strength of spin-dipole mode depicted in fig.\ (\ref{SpinDiQuadrufig})
 is now distributed over a large frequency region. Both the collective and the single particle response is shifted to 
 $\omega>\omega_T$ resulting in a significant 
overlap between the collective mode and the single particle excitations. The collective mode can therefore decay
 into an incoherent motion of individual quasi particles (Landau damping). We also note 
from fig.\ (\ref{SpinDiQuadrufig}) that this overlap with the single particle 
spectrum gives rise to a  second small resonance located at $\omega\simeq1.135\omega_T$. Obviously, such behaviour 
cannot be described within a sum-rule approach. The general shift to a higher frequency for the spin-dipole collective
response as compared to the ideal case ($\omega=\omega_T$) directly follows from the fact that for $a<0$, 
the mean field provides an extra attractive force for the two hyperfine species oscillating in anti-phase and 
the spring constant for this motion is thus enhanced.  

 For the quadrupole mode, there is a very low frequency response in the region
 $0\le\omega\lesssim0.2\omega_T$ as 
is highlighted in the inset in fig.\ (\ref{SpinDiQuadrufig}). This low frequency response is straightforward to understand:
The quadrupole mode is characterized by single particle excitations with angular momentum change $\Delta l=\pm 2$.
The non-interacting spherical harmonic oscillator energies are given by the bands $E_{nl}=(n+3/2)\hbar\omega$ 
with $l=0,2,\ldots n$ for 
 $n$ even and $\l=1,3,\ldots n-1$ for n odd. Hence, the low frequency part of the quadrupole response is simply due to
quasi particle excitations within such an energy band. If the gas was ideal, the frequency of such \emph{intra}-band 
transitions would be $0$ and there would be no net transitions. Due to mean field effects there is a slight dispersion
 within each band~\cite{Bruun1} resulting in a finite  response  at small $\omega$.
For the monopole and spin-dipole modes, we have $\Delta l=0$ and $\Delta l=\pm1$ and there are no intra-band 
transitions excited. These two modes therefore do not have any low frequency response until the interactions are so strong
as to make each harmonic oscillator band overlap such that \emph{inter}-band transitions have a low frequency. 
This low frequency response should have a significant effect on the quadrupole compressibility of the gas since from 
simple 1.\ order perturbation theory one easily sees that the effect of a static quadrupole perturbation on the shape of the 
cloud is determined by the low energy excitations.

The gas is in principle superfluid for $T=0$ for an attractive interaction, and  $k_BT_c\simeq2.8\hbar\omega_T$ 
for the parameters given above~\cite{BCS}. The presence of superfluidity could change significantly 
the low frequency response of the gas and this effect will be examined in a future publication. 
Since the gas strictly is superfluid for $T=0$ for an attractive interaction, we need for consistency 
 to examine whether the effects described above relevant for a gas in the normal phase are still present for $T>T_c$. 
In Fig.\ (\ref{SpinDiQuadrufig}), we therefore  also plot the collective response for $k_BT=3\hbar\omega_T>k_BT_c$
 indicated as a dotted line. 
We see that the effect of a finite temperature on the spectrum is as expected to smear the response out somewhat due to the 
thermal occupation of excited states. Furthermore, as a finite $T$ tends to
 make the gas more dilute and thus diminish the effect of the interactions, the 
spectrum is shifted slightly towards the ideal gas limit. However, the effect is quite small for the low $T$ considered 
here and  the main features of the $T=0$ spectrum remain. 
Importantly, we note that even though $k_BT=3\hbar\omega_T$, the low frequency response ($\omega\lesssim 0.2\omega_T$)
 is not completely suppressed. 
Obviously, for increasing $T$ the response the gas eventually becomes ideal.

\begin{figure}
\epsfig{file=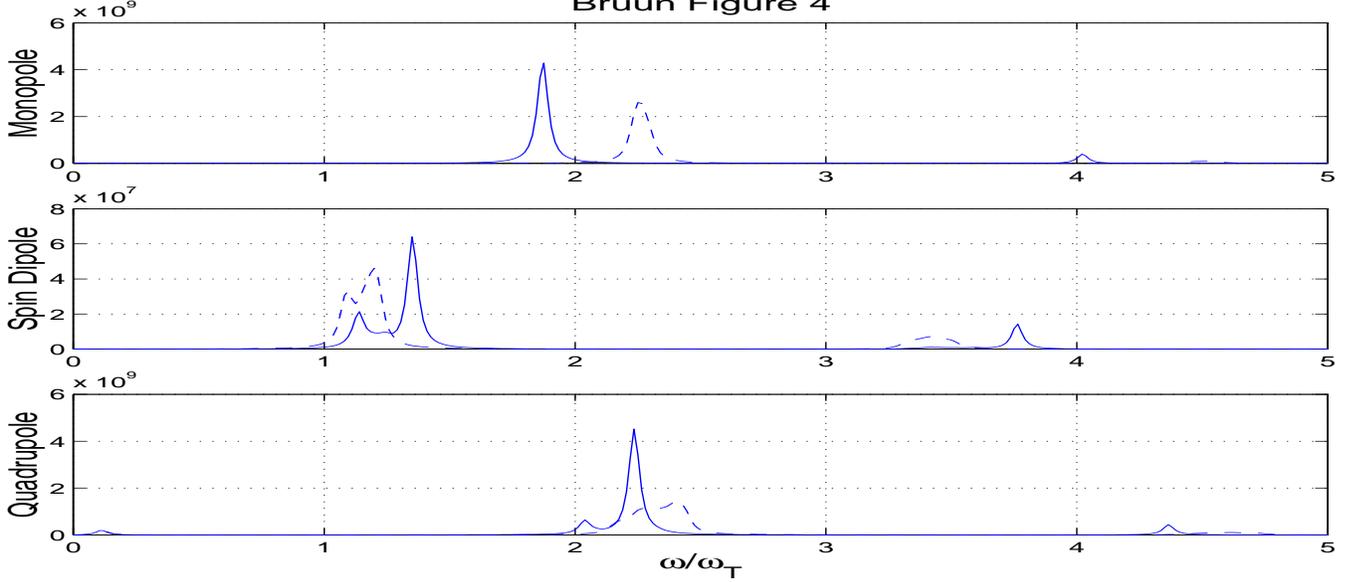,height=0.34\textheight,width=\textwidth,angle=0}
\caption{The $T=0$ collective (solid line) and single particle (dashed line) response
 for the monopole (in units of $l_h^8/\hbar\omega_T$), spin-dipole (in units of $l_h^6/\hbar\omega_T$), and 
quadrupole modes (in units of $l_h^8/\hbar\omega_T$) excited by $F_\sigma(r)=r^4$, 
$F_\sigma(r)=-F_{\bar{\sigma}}(r)=r^3$ and $F_\sigma(r)=r^4$ respectively.}
\label{Highmodes}
\end{figure}

In order to excite higher modes of the gas, one in general needs perturbations containing higher powers of $r$ as can be 
seen if one writes these operators in terms of the raising and lowering 
operators for the harmonic oscillator. This is illustrated  in fig.\ (\ref{Highmodes}), where we plot the response to a
 perturbation given by  $F_\sigma(r)=r^4$, $F_\sigma=-F_{\bar{\sigma}}=r^3$, and $F_\sigma(r)=r^4$
 for the monopole, spin-dipole and quadrupole respectively. We have chosen $\sim 1.6\times10^4$ particles trapped,
 $k_Fa\simeq-0.6$, $\Gamma=0.1\hbar\omega$, and  $T=0$. We see that for such perturbations, higher modes are also excited
as expected. However, the strength of these higher excitations is rather low compared to the lowest modes. This is due 
to the fact that these modes correspond to additional nodes in the density fluctuation profiles and their overlap with the 
perturbing operator $F_\sigma(r)$ is therefore smaller than for the lowest modes. In order to increase the excitation 
rate of these higher modes, one could introduce a perturbation with more such radial nodes. In general, we believe that 
 the excitation of these higher modes is somewhat more complicated than the excitation of the lowest modes, as one needs to
 experimentally generate spatial perturbations of a non-harmonic kind. 
\section{Conclusion}
We have systematically examined the effects of the particle-particle interactions on the collective mode spectrum of a 
normal trapped Fermi gas within a self-consistent RPA scheme. For weak interactions where the 
shift in the eigenfrequencies away from the ideal case is very small, it was shown that a simple sum-rule
 approach to calculate the frequencies of the lowest modes works well. With increasing interaction strength however,
 such a calculation was demonstrated to be complicated by the fact that it is necessary to include the effects of the
 particle-particle scattering on the ground state to achieve accurate results. As the coupling strengths increases
 further, we demonstrated  the emergence of 
effects such as mode splitting and Landau damping coming from the interplay between the collective and the single 
particle excitations.  The importance of such effects was shown to depend on the symmetry 
of the mode. For the quadrupole mode, a very low frequency response was predicted with possible 
implications on the quadrupole compressibility of the gas. The temperature dependence of the modes was considered
 and we finally briefly discussed the excitation of some of the higher collective modes. Due to the possibility of
experimentally manipulating the effective interaction between alkali atoms and the high precision spectroscopy 
typically obtainable in such systems, we believe that some of the effects described in this paper should be 
observable using present day experimental technology. 
\section{Acknowledgements}
We would like to acknowledge valuable discussions with B.\ Mottelson.



\end{document}